 \documentclass[preprint2]{aastex}

\slugcomment{To appear in Publications of the Astronomical Society of the Pacific}

\shorttitle{Spectrum of V532\,Oph at maximum  light}
\shortauthors{Rao et al.}

\begin{document}

%% LaTeX will automatically break titles if they run longer than/n intimated

\title{High-resolution optical spectroscopy of the R Coronae Borealis star V532 
Ophiuchi at maximum light \thanks{Accepted for publication in Publications
of the Astronomical Society of the Pacific.   Based on
observations obtained with the Harlan J. Smith Telescope of McDonald
Observatory of the University of Texas at Austin. } }


\author{N. Kameswara Rao\altaffilmark{2,3},
David L. Lambert\altaffilmark{3},
Vincent M. Woolf\altaffilmark{4}, and
B.P. Hema\altaffilmark{2} }

\altaffiltext{2}{Indian Institute of Astrophysics, Bangalore 560034,
India}
\altaffiltext{3}{The W.J. McDonald Observatory, University of Texas,
Austin, TX 78712-1083, USA}
\altaffiltext{4}{Physics Department, University of Nebraska at Omaha, Omaha, NE 68182-0266, USA}


\begin{abstract}
High-resolution optical spectra of the
R Coronae Borealis (RCB) star V532 Oph at light maximum
are discussed. The absolute visual magnitude M$_V$ of the star is
found to be $-4.9 \pm 0.5$. The  elemental abundances suggest the
star belongs to the majority class of RCB stars but is among the most
O-poor of this class with mild enhancements of heavy elements Y, Zr,
Ba and La.  The C$_{\rm 2}$ Swan  bands are weak in V532 Oph relative
to R CrB.  Other aspects of the high-resolution spectrum confirm
that V532 Oph is representative of majority RCBs, i.e., the radial
velocity is variable, circumstellar material is present and the
photosphere feeds a high-velocity stellar wind.

\end{abstract}




\section{Introduction}

R Coronae Borealis (RCB) stars are supergiants distinguished
by a very H-poor atmosphere and a propensity to decline in brightness
at unpredictable times as clouds of carbon soot intercept the line of
sight to the star. Yet another characteristic of RCB stars is that
they are rare; a catalogue assembled by Jeffery et al. (1996) lists
just 36  `Cool hydrogen-deficient stars', a class which
covers warm and cool RCBs as well as five H-deficient carbon
(HdC) stars which have not exhibited RCB-like declines in brightness.
Comprehensive understanding of the origins and behavior of RCB stars
may depend on increasing the number of known RCB stars and providing
thorough analyses of multi-wavelength observations of new
discoveries. Fortunately, several surveys are indeed increasing
the sample of RCB stars, e.g., Tisserand et al.  (2008, 2011,
2013) and Miller et al. (2012). Clayton (2012) provides a comprehensive 
 list of RCB stars known currently.   

\objectname{V532 Oph}, the subject of this paper, was identified as
a warm RCB star by Clayton et al. (2009)
from ASAS-3 photometry with confirmation provided by low-resolution
optical spectroscopy.
In this paper, our primary goal is to provide an abundance analysis of
V532 Oph and to compare it with other RCBs. This
comparison involves the abundance analysis of 14 warm RCBs by
Asplund et al. (2000) and the analysis by Rao \& Lambert (2003)
of another recent RCB
discovery V2552 Oph (Kazarovets et al. 2003; Hesselbach, Clayton,
\& Smith  2003). The comparison RCB stars like V532 Oph are
all warm RCB with optical spectra dominated by atomic and not
molecular lines. With the exception of Z UMi (Kipper \& Klochkova
2006), cool RCB stars with spectra dominated by C$_2$ and CN lines
have not been subjected to abundance analyses.

Our analysis shows that V532 Oph has a composition representative of
majority RCB stars but stands out with a lower than average O
abundance and heavy element (Y to La) abundances greater than
average. V532 Oph with \objectname{V2552 Oph}, also an O-poor RCB,
hint at a new correlation
between the O abundance and fraction of stellar flux absorbed and
reradiated as the infrared excess emission by the circumstellar dust.

\section{Observations}

Observations of V532 Oph were obtained on two occasions (Table 1) with
the Robert G. Tull cross-dispersed echelle spectrograph of the
Harlan J. Smith 2.7m reflector at the W.J.  McDonald observatory
(Tull et al. 1995).
The spectral resolving power, $R = \lambda/d\lambda $,
employed was 40000. The spectrum covers 3900 to 10000\AA\ with gaps
beyond about 5600\AA\ where the echelle orders were incompletely
captured on the Tektronix 2048 x 2048 CCD.

Two exposures of 30 minutes were combined for each night.
A nearby hot star was observed to remove the telluric line
contribution.  The light curve for the period 2009 to 2014 based on
AAVSO observations shows that the star was at maximum light when
our observations were obtained.

\begin{table*}
\centering
\begin{minipage}{180mm}
\caption{\Large  Spectroscopic Observations of V532 Oph. }
\small\begin{tabular}{llrl}
\hline

Date & Julian Date &   Velocity$^a$   \\
(UT) & (2450000+)  & (km s$^{-1}$) \\
\hline
2010 Apr 21 & 5307.920  & -4.6 $\pm$ 1.5 &      \\
2011 May 16 & 5697.814  & -2.7$\pm$ 0.9 &   \\
\hline
\end{tabular}
\\ $^a$ - Heliocentric radial velocity\\

\label{default}
\end{minipage}
\end{table*}

\section{Interstellar extinction and Absolute magnitude }

In addition to the stellar spectrum, the optical spectrum of V532 Oph shows absorption lines from the
diffuse interstellar medium. This is not surprising since
V532 Oph is located in the Galactic plane.
By using strengths of
interstellar lines to estimate the reddening (and extinction)
and the velocities of the lines with a model of Galactic rotation to estimate the
star's distance, we obtain an estimate of the bolometric magnitude
which acts as a useful constraint in fixing the stellar atmospheric
parameters (see below).

We estimated the
interstellar reddening from the strength of diffuse interstellar
bands (DIBs) on  the high resolution spectra. Fortunately, the
V532 Oph's optical spectrum is a good match
to that of \objectname{R CrB} at maximum
light. R CrB is located at a high galactic latitude and suffers
negligible interstellar reddening including contributions from the
DIBs.  Comparison of the spectra of V532 Oph and R CrB
revealed the DIB features whose strength we could
measure. An example is shown in Figure 1.

\begin{figure}
%\begin{minipage}{120mm}
%\epsfxsize=8truecm
%\epsffile{V532OphismDIB.ps}
\plotone{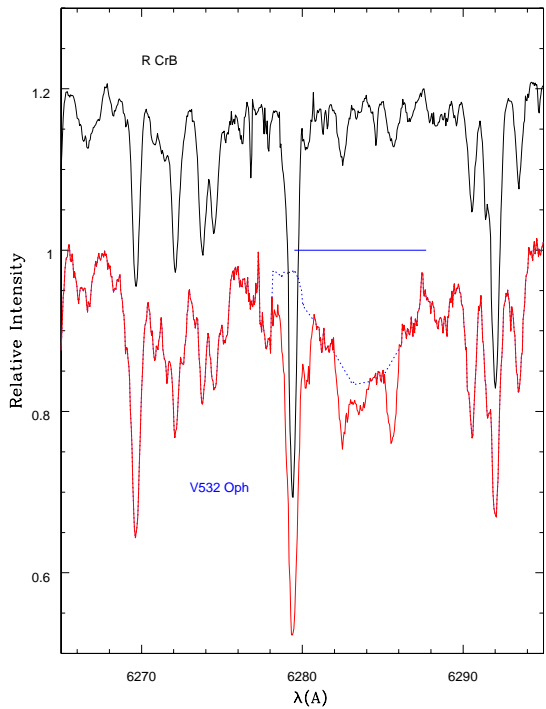}
\caption{Spectra R CrB (top in black) and V532 Oph (bottom in red) are
shown. Both spectra have been corrected for absorption by telluric O$_2$
lines. Note the general similarity of line strengths except for the
broad depression centred at about 6284 \AA\ due to a DIB
along the line of sight to V532 Oph: the dashed blue line shows the estimated
profile of the DIB. }
%\end{minipage}
\end{figure}

\begin{table*}
\centering
\begin{minipage}{180mm}
\caption{\Large Diffuse Interstellar Bands (DIBs) }
\small\begin{tabular}{llll}
\hline

 $\lambda$  & Eq.W & E(B-V) &  \\
\cline{3-4} \\
({\rm \AA})   & ({\rm m\AA}) & Herbig (1993)& Luna et al.(2008)  \\
\hline
 5780  & 316 & 0.64 & 0.69    \\
 5796  & 110 & 0.70 & 0.65    \\
 6284  & 959 & 0.64 & 1.07    \\
 6375  &  28 & .... & ....    \\
 6379  &  75 & 0.83 & 0.85    \\
\hline
\end{tabular}
\label{default}
\end{minipage}
\end{table*}

Equivalent widths of five DIBs were measured (Table 2)  and
the DIB equivalent width -- reddening E(B-V) relations
for four DIBs given by  Herbig (1993) and
Luna et al. (2008) were used to estimate the reddening in Table 2.
The average E(B-V) is 0.76 $\pm$ 0.14 leads to an estimate of
A$_V$ of 2.36$\pm$ 0.43 with R of 3.1. 

The referee of this paper has kindly intimated  that his estimate of
A$_V$ towards V532 Oph based on the calibration of Schlafly \&
 Finkbeiner 
(2011), who use {\it SDSS} spectra and the E(B$-$V) maps of Schlegel et al.
(1998), is 2.43: a value
 very consistent with our estimate.

The light maximum V magnitude has been estimated to be 11.7 by
 Clayton et al. (2009)  and 11.8 by Tisserand et al. (2013). We averaged
 AAVSO's V band magnitudes at maximum obtained since 2010 August to present
 (the star seem to be recovering from a minimum in the earlier period)
 as 11.46 $\pm$ 0.15 (95 measurements). This V maximum value
may then be corrected for interstellar
extinction resulting in V$_0$ of 9.10 $\pm$ 0.45.

\begin{figure}
%\epsfxsize=8truecm
%\epsffile{V532Oph11m16NaDvlsr.ps}
\plotone{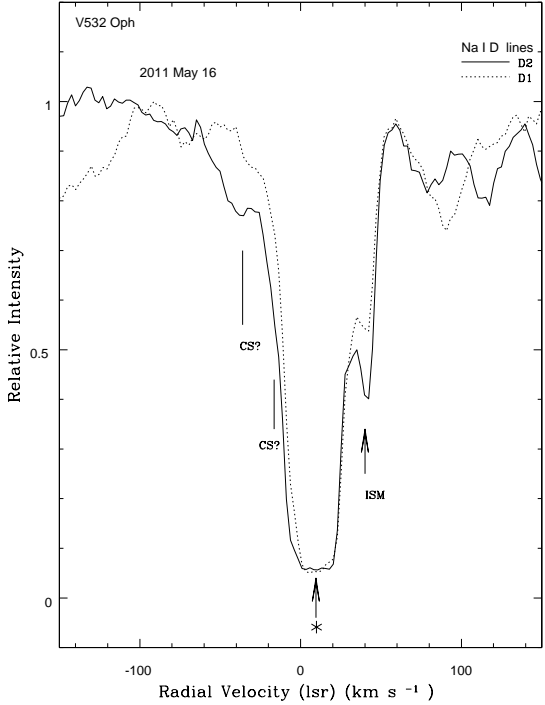}
\caption{NaI \,{\sc i} D$_2$ (full line) and D$_1$(dashed line) lines
in the spectrum of V532 Oph are shown plotted with respect to the
LSR velocity.
The strongest line is attributed to the star. A red-shifted component
at 40.5 km s$^{-1}$
is identified with the interstellar medium and labelled ISM. Two
features to the
blue of the stellar line are tentatively attributed to the
circumstellar gas and labelled  CS?.}
\end{figure}

Kinematics of the interstellar gas along the line of sight to V532 Oph
are best revealed by the Na D lines, even though interstellar
components must be distinguished from circumstellar components.
Figure 2 shows the Na D profiles where D$_1$ and D$_2$ profiles are
superimposed. (The oscillator
strength of the D$_2$ line is twice that of the D$_1$ line).
As an aid to identifying the interstellar Na D components, we computed
the expected radial velocity with respect to the LSR with distance in
the direction of V532 Oph using the model
Galactic rotation  given by Brand \& Blitz (1993).
Components with negative LSR radial velocities are not expected to
occur in this direction. However,  components
are possibly present in the Na\,{\sc i} D profiles of
V532 Oph at -35 and -17 km s$^{-1}$
which we attribute to  outward-flowing circumstellar gas.
There is a prominent absorption component to both Na\,{\sc i} D lines
at the LSR radial velocity
of 40.5 km s$^{-1}$ (marked as ISM in Figure 2).

The Galactic rotation curve
suggests such a radial velocity
occurs for a star  at a distance of 6.17 kpc. The presence of
such a  cloud in front of a star provides a minimum
distance to the star. This distance estimate provides an independent
means of arriving at a lower limit to
the absolute visual magnitude (M$_V$) of this star.
The extinction-corrected V magnitude 9.1 obtained earlier coupled
with the distance estimate of 6.17 kpc
provides the M$_V$ of -4.85 $\pm$ 0.5
consistent with estimates obtained from warm
RCB stars in LMC (Tisserand et al. 2009). The bolometric
correction for V532 Oph (T$_{\rm eff}$ of 6750 K -see later) is
expected to be
small:  $\sim$ 0.019  according to Flower (1996) for normal
supergiants.  Thus,
M$_{bol}$ is expected to be about -4.9 $\pm$ 0.5 as closely anticipated by 
 Tisserand et al. (2009) for RCBs with mean $T_{\rm eff}$ 6750 K and a V-I  color of 0.6.

\begin{figure}
%\epsfxsize=8truecm
%\epsffile{V532OphloggTeffN.ps}
\plotone{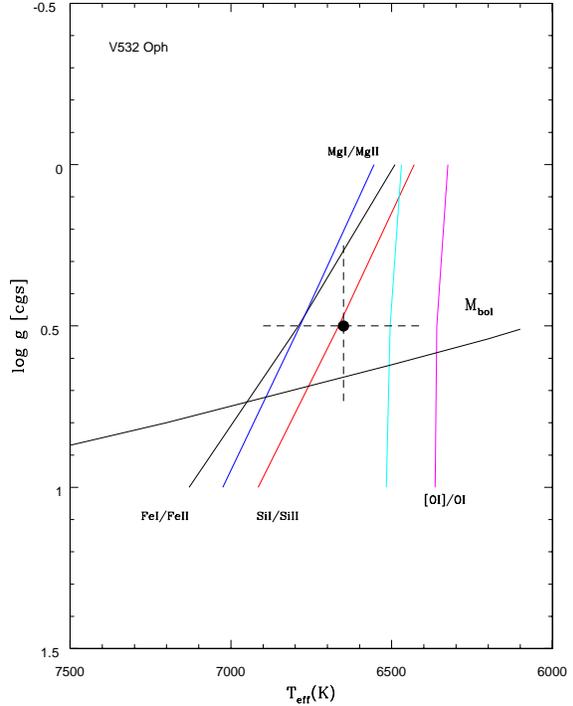}
\caption{Loci of ionization equilibria and other parameters in the
$T_{\rm eff}$ -- $\log g$
plane from the spectrum on 2011 May 16. The $M_{\rm bol}$ relation
suggested by Asplund et al. (2000) is shown as full line
(see $M_{\rm bol})$. The relation
using the $M_{\rm bol}$ estimated for V532 Oph in this paper is 
same
as that given by Asplund et al. (2000).
Two sigma error in the equivalent estimate of 
the [O\,{\sc i}] line could push the locus of [O\,{\sc i}]/O\,{\sc i}
to higher $T_{\rm eff}$ (shown in cyan). Final choices of
$T_{\rm eff}$ and $\log g$ are 
indicated by a large dot. The error range in
these values is shown by dashed lines. }
\end{figure}

\section{Abundance analysis}

The abundance analysis of V532 Oph followed procedures developed
and applied by
Asplund et al. (2000) to a sample of 14 warm RCBs.
The adopted line-blanketed hydrogen-deficient
model atmospheres  are described by Asplund et al. (1997).
An abundance ratio C/He = 1\% by number
of atoms was assumed.
As discussed thoroughly by Asplund et al. (2000),
the observed C\,{\sc i} lines in
these RCBs (also, here in V532 Oph) do not
return the input C abundance but a value about 0.6 dex lower. This
inconsistency referred to by Asplund et al.  as `the carbon problem'
has yet to be explained in a self-consistent manner (i.e., the
input C abundance for model atmosphere construction reproduces the
observed carbon spectroscopic features e.g., C\,{\sc i}, C\,{\sc ii}
and C$_2$ lines) with their known $gf$-values. Asplund et al. suggest 
that relative abundance such as O/Fe may be insensitive to the correct
resolution of the carbon problem.

Choice of the appropriate model atmosphere requires selection of
effective temperature (T$_{\rm eff}$), surface gravity ($g$),
microturbulence ($\xi$), and the C/He ratio (here = 1\%). Since the
T$_{\rm eff}$ and $g$ determinations are somewhat dependent on
$\xi$, we began with the latter's determination. For all species
represented by lines spanning a range in equivalent widths, we
determined the value of $\xi$ for which the standard deviation
of derived abundances was at a minimum. This minimum
was found for a similar value of $\xi$ for a majority of the
species from N\,{\sc i} to Y\,{\sc ii}, i.e., $\xi = 7.5$ km s$^{-1}$.

Several conditions are used to locate loci in
the T$_{\rm eff} - \log g$ plane.
Ionization equilibrium was demanded using   Fe\,{\sc i}/Fe\,{\sc ii},
Mg\,{\sc i}/Mg\,{\sc ii},  and Si\,{\sc i}/Si\,{\sc ii}
to  provide loci in the $T_{\rm eff}$ - log $g$ plane.
Excitation equilibrium  through the use of  [O\,{\sc i}] and
high-excitation O\,{\sc i} lines  provides a gravity-insensitive
temperature indicator. We also used the recipe
suggested by Asplund et al. (2000) which adopts a
$M_{\rm bol}$ of -5.0 for  R CrBs, to
obtain a relation between $T_{\rm eff}$ and log $g$.  The
adopted parameters are indicated in
Figure 3 along with the loci of various
indicators in the $T_{\rm eff}$ -- $\log g$ plane.

The stellar parameters $T_{\rm eff}$ = 6750$\pm$250 K,
$\log g$ = 0.5$\pm$0.3 (cgs units)
and microturbulence $\xi$$_{\rm tur}$ = 7.5$\pm$1.0 km s$^{-1}$
were chosen. The derived
abundances are given in Table 3 along with the abundances of
R CrB from Rao \& Lambert (2003) and the dependence of V532 Oph's
abundances on a 250 K change in effective
temperature and a 0.5 dex change in $\log g$ (sampling intervals in our
 atmospheric models) .
% Differential abundances with respect to the Sun (Lodders 2003)
%are given as [El/Fe] in Table 3.
The final column headed $\Delta$ gives the difference between
the logarithmic abundances of V532 Oph and R CrB.

By chemical composition, V532 Oph is a
representative member of the majority
class of RCB stars (Lambert \& Rao 1994)
and is set clearly apart from the
minority RCBs with their low Fe abundance and very high Si/Fe and S/Fe
abundance ratios. Within the majority class, there
are real differences in the
abundances of certain elements (e.g., H, Li, O and Y)
and very little differences
in abundance ratios within the elements from Na to Ni.
V532 Oph's composition reflects these differences.
For the  elements Na to Ni, the composition of V532 Oph is remarkably
similar to that of R CrB -- entries in the column headed $\Delta$
range from +0.16 to $-0.13$ for these elements where
the differences are comparable
or even less than the measurement uncertainties.
In contrast, the differences for H, Li,
O and Y are much larger than these uncertainties. That these are real
differences is directly shown by direct comparison of spectra.
For example, the lower O abundance of V532 Oph relative to R CrB is
well shown in Figures 4 and 5. Figure 4
compares spectra of R CrB, V532 Oph and V2552 Oph
(coincidentally, a similarly O-poor RCB - Rao \& Lambert 2003)
around the 6156 \AA\ O\,{\sc i} multiplet.
Figure 5 compares spectra of the same three stars around the
6363 \AA\ [O\,{\sc i}] line.
(The stronger forbidden line at 6300 \AA\ is not
on the recorded portion of the spectra.) Finally, Figure 6 extends the
comparison to include a Y\,{\sc ii} line which relative to its
strength in R CrB is enhanced in both V532 Oph and V2552 Oph.
These three RCBs have similar atmospheric parameters and,
therefore, differences
in the strength of a line are  a reflection of an abundance
difference.

One additional spectroscopic difference between V532 Oph
(and V2552 Oph) and R CrB deserves illustration, namely the weakness
of the C$_2$ Swan bands in
the former pair relative to R CrB -- see Figure 7.
This difference was noted earlier
by Clayton et al. (2009).
The weaker Swan bands for V532 Oph presumably reflects
the lower C abundance where the Swan band strength depends
on the square of the
C abundance and a 0.2 dex difference (Table 3) corresponds to
nearly a factor of three
decrease in the Swan band strength.

Although these samples of spectra of R CrB, V2552 Oph and V532 Oph
serve to highlight differences between R CrB and the latter two
stars, it should be pointed out that all three stars are fair
representatives of `majority' RCBs among which, as noted above,
there is an intrinsic
spread in abundance for O and Y elements among a few others.
In Figure 8, we show histograms for O, S, Fe and Y abundances
for majority and minority RCBs with R CrB, V532 Oph and V2552 Oph's
location in the histograms identified.  Figure 8
clearly shows that  R CrB is among the O-richest and
V532 Oph with V2552 Oph are among the
O-poorest majority RCBs.
In contrast, the S histogram is much narrower and R CrB,
V532 Oph and V2552 Oph are in the central core of the histogram.
Finally, the histogram for Y shows a real 2 dex spread in Y
abundances with R CrB with one of the lowest Y
abundances and V532 Oph and V2552 Oph
close to the mean for majority RCBs.

\begin{table*}
\centering
\begin{minipage}{180mm}
\caption{ \Large
  Elemental Abundances for V532 Oph}
\small\begin{tabular}{lcllrl}
\hline\hline
 & V532 Oph  &n$^{a}$  & \underline{$\delta$ $T_{\rm eff}$, $\delta$ $\log g$ } &
R CrB$^{b}$  & $\Delta$$^{c}$
\\
\  Species  &     &  & 250 K, 0.5 & \\
\hline
H\,{\sc i} & 6.31  & 1 & 0.14, 0.12 & 6.86 & -0.55 \\
Li\,{\sc i} &$<$0.97  & 1  &   ....  & 2.55 & $<$-1.6\\
C\,{\sc i} & 8.91$\pm$ 0.34 & 14 & 0.01, 0.02 & 9.11 &0.20 \\
N\,{\sc i} & 8.57$\pm$0.23 & 10 & 0.17, 0.15  & 8.42  & +0.15 \\
O\,{\sc i} & 7.90$\pm$0.34 & 6 & 0.11, 0.14   & 8.60  &-0.70 \\
Ne\,{\sc i}& 8.62$\pm$0.21 & 3 & 0.35, 0.30   & ....  & ....     \\
Na\,{\sc i} & 6.22$\pm$0.04 & 4 & 0.17, 0.15  & 6.11  & 0.11 \\
Mg\,{\sc i} & 6.84$\pm$ 0.10 & 3 & 0.17, 0.15 & 6.81  &0.03 \\
Mg\,{\sc ii} & 6.89    & 1       & 0.09, 0.10 & ....      & .... \\
Al\,{\sc i} & 5.81$\pm$ 0.15 & 4 & 0.15, 0.15 & 5.76  & 0.05 \\
Si\,{\sc i} & 6.93$\pm$ 0.21 & 7 & 0.16, 0.15 & 6.97  & -0.04 \\
Si\,{\sc ii} &6.82    & 1  & 0.18, 0.18       & ....  & .... \\
S\,{\sc i} & 6.76$\pm$ 0.23 & 7 & 0.08, 0.07  & 6.70  & 0.06 \\
K\,{\sc i} & 4.83    & 1      & 0.25, 0.19    & 4.77  & 0.06 \\
Ca\,{\sc i} & 5.19$\pm$ 0.21 & 6 & 0.22, 0.19 & 5.32  &-0.13 \\
Sc\,{\sc ii} & 2.80$\pm$ 0.27 & 3 & 0.10,0.08 & 2.89  &-0.09 \\
Ti\,{\sc ii} & 4.21$\pm$ 0.26 & 2 & 0.05, 0.11& 4.05  & 0.16 \\
Fe\,{\sc i} & 6.47$\pm$ 0.21 & 31 & 0.21,0.15 & 6.40  & 0.07\\
Fe\,{\sc ii} & 6.51$\pm$ 0.13& 10 & 0.00, 0.12& 6.40  & 0.11 \\
Ni\,{\sc i} & 5.55$\pm$ 0.13 & 5 & 0.22, 0.15 & 5.49  & 0.06 \\
Cu\,{\sc i} & 4.08$\pm$ 0.20 & 2 & 0.18, 0.14 & ....  & .... \\
Zn\,{\sc i} & 4.41 &  1          & 0.17, 0.14 & 4.16  &0.25 \\
Y\,{\sc ii} & 2.08$\pm$ 0.09 & 5 & 0.10, 0.08 & 1.55  & 0.53 \\
Zr\,{\sc ii} & 2.11$\pm$ 0.15& 3 & 0.08, 0.10 & 1.84  & 0.27 \\
Ba\,{\sc ii} & 1.45$\pm$ 0.05 & 5 & 0.14,0.07 & 1.13  & 0.32 \\
La\,{\sc ii} & 0.67$\pm$ 0.30 & 2 & 0.17, 0.03& 0.64  &0.03 \\
\hline\hline
\end{tabular}
\\ $^{a}$ n = number of lines used in the analyses.
\\ $^{b}$ From Rao \& Lambert (2003).
\\ $^{c}$ Abundance difference between V532Oph and R CrB.

\label{default}
\end{minipage}
\end{table*}

\section{V532 Oph is a typical RCB: wind and dust}

Other aspects of V532 Oph's optical spectrum at maximum light
show that the star is a typical majority RCB. In particular,
there is evidence for a wind off the photosphere. This is clearly
shown by the extended blue wing to the O\,{\sc i} 7771 \AA\ line
(Figure 9) and other strong lines.
The implied terminal velocity indicated by each strong line correlates
well with the lower excitation potential (LEP) of the line.
The velocity-LEP
relation is very similar to that shown for R CrB (Rao, Lambert
\& Shetrone 2006).

\begin{figure}
%\epsfxsize=8truecm
%\epsffile{V532OphOI6155.ps}
\plotone{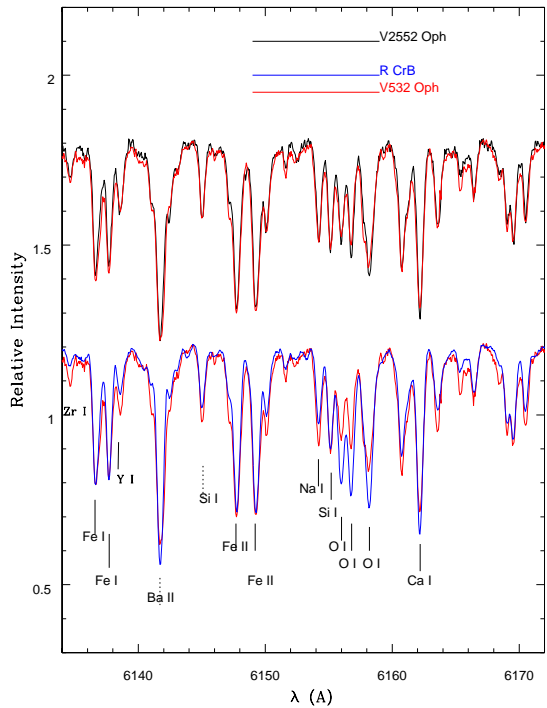}
\caption{Comparison of the spectrum of V532 Oph in the O\,{\sc i}
6155\AA\ region with V2552 Oph (black line)  and R CrB
(blue line). Note the weakness
of O\,{\sc i} lines in both V532 Oph (red line) and V2552 Oph relative
 to their
strength in R CrB.}
\end{figure}

\begin{figure}
%\epsfxsize=8truecm
%\epsffile{V532OphSiII6371.ps}
\plotone{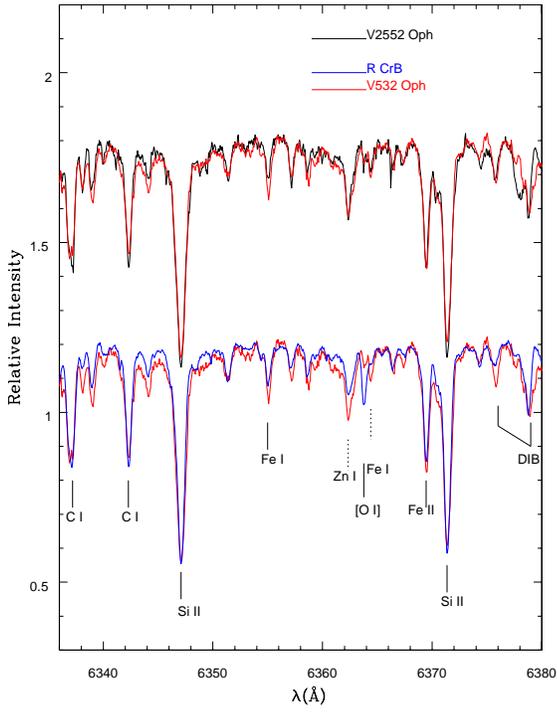}
\caption{Comparison of the  of [O\,{\sc i}] 6363\AA\ line in the
spectra of V532 Oph (red line),   V2552 Oph
(black line) and R CrB (blue line).
Note the weakness of [O\,{\sc i}] 6363\AA\
in the spectra of V532 Oph and V2552 Oph relative to the spectrum of
R CrB (blue line).}
\end{figure}

\begin{figure}
%\epsfxsize=8truecm
%\epsffile{V532OphYII7450.ps}
\plotone{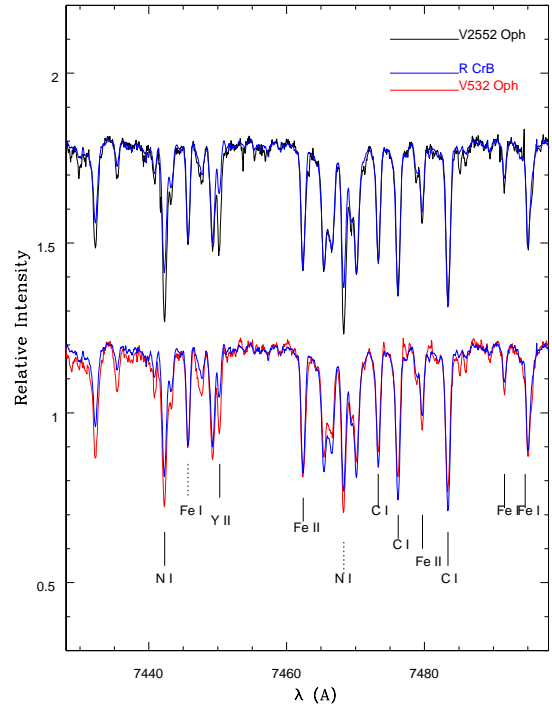}
\caption{Comparison of the strengths of a Y\,{\sc ii} line
in the V532 Oph spectrum
(red line),the  V2552 Oph spectrum (black line) as well as R CrB (blue
line).}
\end{figure}

\begin{figure}
%\epsfxsize=8truecm
%\epsffile{V532Ophc25635.ps}
\plotone{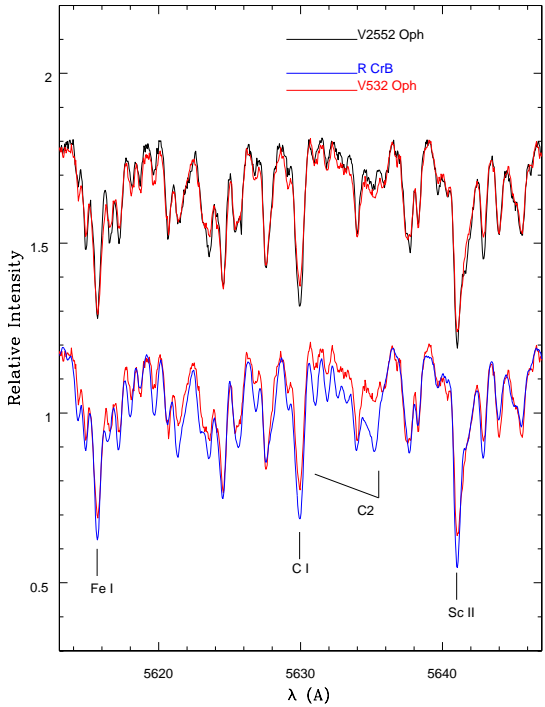}
\caption{The weakness of C$_2$ bands in V532 Oph (and V2552 Oph) is
shown.  The Swan blue-degraded 0-1
bandhead at 5635 \AA\ region is shown for V532 Oph
(red line), V2552 Oph (black line) and R CrB (blue line).
The C$_2$ band in both V532 Oph and V2552 Oph is weaker than in R CrB.}
\end{figure}

\begin{figure}
%\epsfxsize=8truecm
%\epsffile{HistOYFe.ps}
\plotone{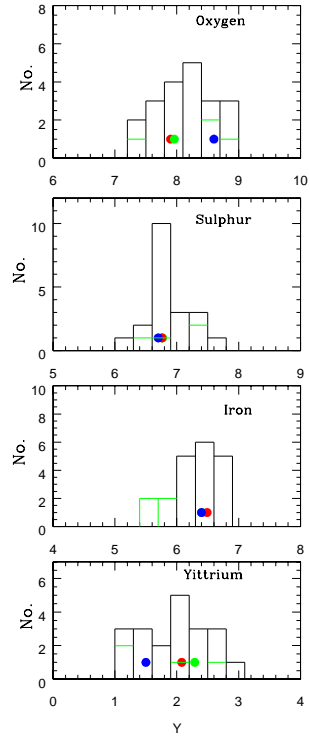}
\caption{ Histograms for elemental abundances of
oxygen, sulphur, iron and yttrium for warm  RCBs.
The green line  refers to the four known minority RCBs.
The black line refers to the total  RCBs in the bin (abundance) , with
a red dot denoting the position of V532 Oph,
the green dot the position
of V2552 Oph and a blue dot the position of R CrB. 
Abundances are from Asplund et al. (2000), 
Rao \& Lambert (2003, 2008) and this paper. }
\end{figure}

\begin{figure}
%\epsfxsize=8truecm
%\epsffile{V532OphOI7771v.ps}
\plotone{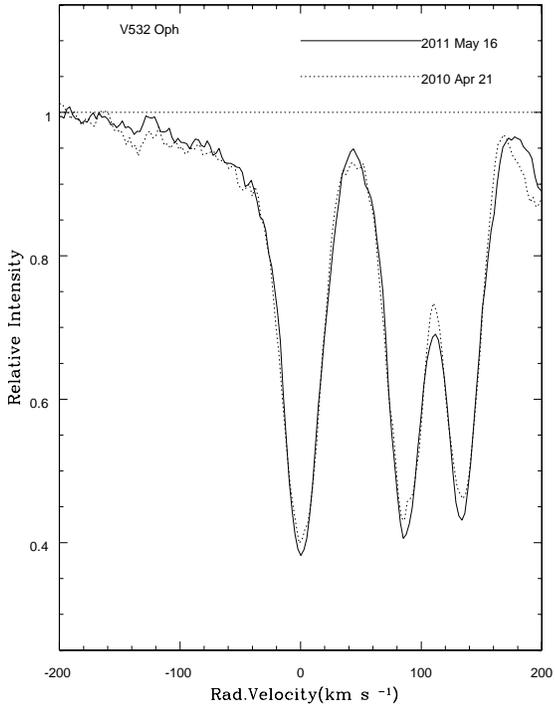}
\caption{The blue extension to the O\,{\sc i} 7771 \AA \, line
to a velocity of
-125 $\pm$5 km s$^{-1}$ suggests a stellar wind may be operating continuously.}
\end{figure}

\begin{figure}
%\epsfxsize=8truecm
%\epsffile{WISEV532Ophsed1.ps}
\plotone{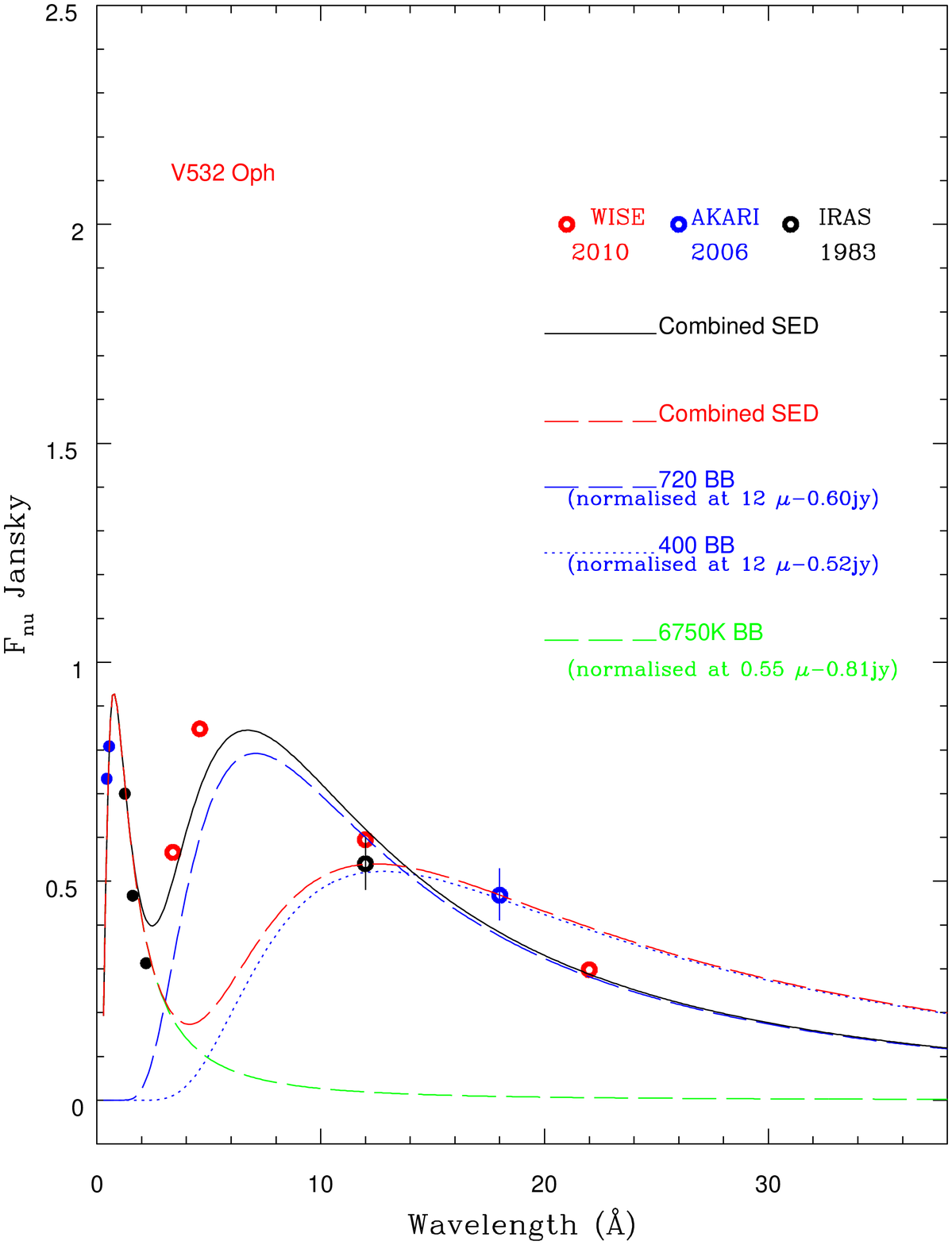}
\caption{ The dereddened spectral energy distribution
fit with blackbodies of 6750 K for the  star and 400K (IRAS
and AKARI) and 720K (WISE) for dust .
The  dust emission was similar  in 1983 (IRAS)
and AKARI (2006) but in 2010 (WISE)  it was hotter. }
\end{figure}

%\subsection{Infrared excess from circumstellar dust}

A feature of RCBs is their ability to form carbon soot.
If a soot cloud
intercepts the line of sight, it causes the characteristic dimming of
the star. Such a cloud and others existing off the line of sight
absorb starlight and emit in the infrared to provide an infrared
excess. Thanks to clouds off the line of sight,
an infrared excess will be present
even in the absence of a cloud along the line of sight, i.e.,
when the star is at maximum light.
Warm dust may contribute excess flux even at the K or even the
H band.  Clayton et al. (2009) noted that
the interstellar reddening-corrected  J-H , H-K colours of V532 Oph
appeared unaffected by dust emission.
Cooler dust gives an excess at longer wavelengths.
For V532 Oph,  mid-infrared measurements from the AKARI satellite
(Ishihara et al. 2010) detected excess emission
at 18$\mu$m. IRAS also detected the source at
12$\mu$m but not at 25$\mu$m (Helou \& Walker 1988).

Our recent
investigation of {\it Spitzer} spectra explored the characteristics of
infrared excesses for a large sample of RCBs (Garc\'{\i}a-Hernandez
et al. 2011, 2013). In this investigation, a
spectral energy distribution (SED) for a star  was constructed from
optical  and infrared photometry corrected for interstellar reddening.
A SED for V532 Oph was constructed for E(B-V) = 0.76
with optical photometry from  Clayton et al. (2009) and
near-infrared photometry from 2MASS (Cutri et al. 2003).
Mid-infrared photometry is from satellites IRAS in 1983, AKARI in 2006
and WISE (Wright et al. 2010) in 2010 (Figure 10).
The infrared flux in 1983 and 2006 was similar in
strength but was appreciably stronger in 2010. A blackbody fit to the
stellar fluxes and the infrared emission by dust gives a stellar
blackbody temperature of 6750 K and dust temperatures of 400 K
in 1983 and 2006 and 700 K in 2010.
The fraction of stellar flux absorbed and
reemitted by the dust is R = 0.03 in 1983 and
2006 but R = 0.09 in 2010.
This value of R is among the lowest values for RCB stars. It is
intriguing that V532 Oph has a low O abundance and a low value of
R.  In a subsequent
paper, we explore the possibility that the
amount of circumstellar dust
as measured by the fraction R is correlated with the O abundance.

\section{Concluding remarks}

Our high-resolution spectroscopic analysis of the recently
discovered RCB V532 Oph (Clayton et al. 2009) confirms that it is
a warm RCB. It belongs to the majority RCB class (Lambert \& Rao 1994)
with a lower than average O abundance and an average Y abundance.
It exhibits
a stellar wind and circumstellar gas and dust. The dust content as
sampled in 1983, 2006 and 2010 is less than average for the RCB sample
studied by Garc\'{\i}a-Hern\'{a}ndez et al. (2011).

\acknowledgments
   We would like to express our thanks to the anonymous referee for pointing out
 an error as well as for his very useful comments. 
We acknowledge with thanks the variable star observations from the
AAVSO international data base contributed by world
wide observers used in this research.
This research has made use of the SIMBAD database, operated
at CDS, Strasbourg, France.
DLL acknowledges support from the Robert A. Welch Foundation
of Houston, Texas through grant F-634.
NKR would like to thank Instituto de Astrofisica de
Canarias, Tenerife and Arturo Manchado,
Anibal Garc\'{\i}a-Hern\'{a}ndez for inviting him as
a Severo Ochoa visitor during January to April 2014 when part of this work was done.

%% in parentheses, after the keyword, but they will not be verified.

{\it Facilities:} \facility{McD: Smith}.

%\begin{thebibliography}{}

%\end{thebibliography}

\end{document}